\documentclass{emulateapj}
\usepackage{natbib,graphicx,amsmath,amsthm,ulem,color}

\newcommand{\be}{\begin{equation}}
\newcommand{\ee}{\end{equation}}
\newcommand{\beqn}{\begin{eqnarray}}
\newcommand{\eeqn}{\end{eqnarray}}
\newcommand{\bi}{\begin{itemize}}
\newcommand{\ei}{\end{itemize}}

\newcommand{\E}{Z_{\rm free}}

\def\refnew#1{(\ref{#1})}

\def\pomega{\varpi}

\def\cm{\, \rm cm}

\def\g{\rm g}
\def\pomega{\varpi}
\def\deg{^\circ}

\begin{document}

\title{Extracting Planet Mass and Eccentricity From TTV data }
\author{Yoram Lithwick\altaffilmark{1},   Jiwei Xie\altaffilmark{2}, \& Yanqin Wu\altaffilmark{2} }
\altaffiltext{1}{Dept. of Physics and Astronomy, Northwestern University, 2145 Sheridan Rd., Evanston, IL 60208
\& Center for Interdisciplinary Exploration and Research in Astrophysics (CIERA)}
 \altaffiltext{2}{Department of Astronomy and Astrophysics, University of Toronto, Toronto, ON M5S 3H4, Canada}

\begin{abstract}
  Most planet pairs in the Kepler data that have measured transit time
  variations (TTV) are near first-order mean-motion resonances. We
  derive analytical formulae for their TTV signals. We separate planet
  eccentricity into free and forced parts, where the forced part is
  purely due to the planets' proximity to resonance.  This separation
  yields simple analytical formulae.  The phase of the TTV depends
  sensitively on the presence of free eccentricity: if the free
  eccentricity vanishes, the TTV will be in phase with the longitude
  of conjunctions. This effect is easily detectable in current TTV
  data.  The amplitude of the TTV depends on planet mass and 
  free eccentricity, and it determines planet mass uniquely only when
  the free eccentricity is sufficiently small. We proceed to analyze
  the TTV signals of six short period Kepler pairs.   We find that
    three of these pairs (Kepler-18,24,25) have TTV phase consistent
    with zero.  The other three (Kepler-23,28,32) have small TTV
    phases, but ones that are distinctly non-zero.  We deduce that the
    free eccentricities of the planets are small, $\lesssim 0.01$, but
    not always vanishing.  Furthermore, as a consequence of this, we
    deduce that the true masses of the planets are fairly accurately
    determined by the TTV amplitudes, within a factor $\lesssim 2$.
    The smallness of the free eccentricities suggests that the planets
    have experienced substantial dissipation.  This is consistent with
    the hypothesis that the observed pile-up of Kepler pairs near
    mean-motion resonances is caused by resonant repulsion.  But the
    fact that some of the planets have non-vanishing free eccentricity
    suggests that after resonant repulsion occurred there was a
    subsequent phase in the planets' evolution when their
    eccentricities were modestly excited, perhaps by interplanetary
    interactions.   
\end{abstract}

\section{Introduction}
\label{sec:intro}

The Kepler mission has detected an abundance of low-mass close-in
planets \citep{Batalhaetal12}.  Remarkably, hundreds of them are
members of planetary systems \citep{Lissaueretal11,Fabryckyetal12}.
These will likely prove to be a Rosetta stone
for deciphering the dynamical history of planetary systems.

One of the most intriguing Kepler discoveries is that, while the
spacing between planets in a system appears to be roughly random,
there is a distinct pile up of planet pairs just wide of certain
resonances, and a nearly empty gap just narrow of them
\citep{Lissaueretal11,Fabryckyetal12}. In \citet{lithwua}, we proposed
that dissipation is responsible for this asymmetry, via an effect we
termed ``resonant repulsion'' \citep[also see the
independent work by][]{baty}.   The eccentricity of planets near resonances can be
separated into two parts: a part that is forced by the resonance and
is determined by the planets' proximity to resonance ({\it forced
  eccentricity}), and a part that is unrelated to the resonance ({\it
  free eccentricity})\footnote{Our forced eccentricity is perhaps more
  accurately called the forced resonant eccentricity to distinguish it
  from the more commonly used forced {\it secular} eccentricity
  \citep[e.g.,][]{MD00}. But only the resonant contribution plays a
  role in this paper.}. If there is dissipation, it damps away the
planets' free eccentricities, but the forced eccentricities persist as
long as the planets remain close to resonance. As the dissipation
continually acts on these forced eccentricities, it extracts energy
from the planets' orbits, and in doing so pushes apart any planet pair
that happens to lie near a resonance \citep{lwpluto,paprr}. Hence all
such planet pairs end up just wide of resonance, naturally explaining
the Kepler result.

If it is indeed resonant repulsion that is responsible for the pile up
-- and to date no other tenable mechanisms have been proposed -- then
there are a number of interesting implications. 
 First, it implies that before resonant repulsion occurred the
 distribution of spacings was nearly uniform, i.e., that planet pairs
 were placed with little regard for resonances. 
 Second, it implies that most of the Kepler planets suffered a
 prolonged bout of eccentricity damping. 
 Third, it implies that the free eccentricities of the planets should
 be zero today, after the prolonged bout of dissipation.  This
 prediction can be tested using the transit time variations (TTV)
 recorded by Kepler, as we demonstrate in this paper.

 A transiting planet that has no companion transits at perfectly
 periodic times.  But one that has a companion deviates slightly from
 its periodic schedule because of the gravitational tugs from its
 companion.  \cite{Agol} and \cite{HolmanMurray} proposed using TTV
 signals to characterize the companions of transiting extrasolar
 planets, and this technique has proved to be highly successful both
 for confirming Kepler candidates and for measuring their masses and
 eccentricities
 \citep[e.g.,][]{CochranKepler18,Ford12,SteffenIII,FabryckyIV}.
 However, all these studies rely on fitting the observed TTV signals
 to direct N-body simulations \citep[e.g.][]{verasford}.  Such fits
 are computationally costly.  Moreover, N-body simulations do not
 provide a dynamically transparent interpretation of the system.

Here, we focus on near-resonant pairs because the nearly coherent
interactions in such pairs induce particularly large TTV signals. Such
pairs account for most of the TTV detections to date in the Kepler
database.  Motivated by our earlier work on resonant repulsion, we
separate the eccentricity into free and forced parts.  Interestingly,
in so doing, the expression for the TTV near first-order resonance
becomes particularly simple.

This paper is organized as follows.  In Section
\ref{sec:planetparams}, we present new analytical formulae for the TTV
from two near-resonant planets, and show that the results agree
with N-body simulations.  In Section \ref{sec:data}, we apply the TTV
formulae to six planet pairs with published TTV data.  
In Section \ref{sec:sum}, we discuss our findings and their
implications.

\section{Planet Parameters from  Analytical TTV}
\label{sec:planetparams}

We consider the TTV signals from two coplanar planets that lie near
(but not in) a $j\!\!:\!\!j\!-\!1$ mean motion resonance.  The results are derived
in the Appendix, and summarized in the following. 
 Let $\delta t\equiv O-C$
 be the
inner planet's transit time delay, where $O$ is the observed
transit time and $C$ is 
 calculated from the linear ephemeris
under the assumption that transits are perfectly periodic.
Similarly, $\delta t'$ is  the outer planet's time delay. We show in the Appendix that
\beqn
\delta t&=&{V\over 2i}e^{i\lambda^j}+c.c.  =|V|\sin(\lambda^j+\angle
V)
\label{eq:tv}  \\
\delta t'&=&{V'\over 2i}e^{i\lambda^j}+c.c.
=|V'|\sin(\lambda^j+\angle V')
\label{eq:tvp} \ , \eeqn 
where $V= |V|e^{i\angle V}$ and $V'= |V'|e^{i\angle V'}$ are the complex TTV
(expression in Eqs. \ref{eq:v}--\ref{eq:vp}), and are nearly constant; 
 c.c. denotes the complex conjugate of the
preceding term; and 
  \be \lambda^j\equiv
  j\lambda'-(j-1)\lambda\, , \label{eq:lamj} 
  \ee 
  is the
    {\it longitude of conjunctions} \citep[e.g.,][]{Agol},
where
 $\lambda$ and
  $\lambda'$ are the mean longitudes of the inner and outer planet,
  respectively.
  At the order of approximation to which we work  (see Appendix), we may
set
  \beqn \lambda={2\pi\over P}(t-T) , \ \
  \lambda'={2\pi \over P'}(t-T') \label{eq:lams} 
  \eeqn 
  in Equation (\ref{eq:lamj}), 
  where
  $P$ and
  $P'$ are the periods of the inner and outer planet, and $T$ and $T'$
  are offsets; all four of these parameters are constant.
   The actual mean longitudes differ slightly from the above expressions if the
  planets are eccentric (and that  is included in the TTV derivation).

  Equations \refnew{eq:tv}--\refnew{eq:tvp} show that TTV signals are
  sinusoidal, with a period determined by $\lambda^j$. We call this
  the {\it super-period}:
\begin{equation}
  P^j\equiv {1\over |{j/P'-(j-1)/P|} }  \ .
    \label{eq:psuper}
  \end{equation}
  There is 
 a simple geometrical
  interpretation to $\lambda^j$.
    Since $\lambda$ and $\lambda'$ are approximately
  the angular positions of the inner and outer planets,
    $\lambda^j$ gives the angular position of
  both planets whenever they hit conjunction ($\lambda =
  \lambda'$).  
  Furthermore, $\lambda^j$
   progresses
  linearly in time between conjunctions.  
  For a near-resonant pair, successive conjunctions differ only slightly in 
  angular position, and hence their super-period is very long.
  More precisely, 
  we
 define the normalized distance to resonance
  \be
  \Delta\equiv {P'\over P }{j-1\over j}-1  \  , 
  \label{eq:Delta}
  \ee
  in which case the super-period is
 \begin{equation}
  P^j
= {P'\over j|\Delta|}  \ .
    \label{eq:psuper2}
  \end{equation}
  For example, a planet pair that has a period ratio of $2.02$ is at a
  distance of $\Delta = 0.01$ from the 2:1 resonance, and its
  super-period is 50 times longer than the outer planet's orbital
  period.  A pair of planets that have $\Delta>0$ lie
  wide of resonance, and their $\lambda^j$ decreases with time (i.e.,
  is retrograde with respect to the orbital motion); conversely, a
  pair with $\Delta<0$
  lie
   narrow of resonance and their
  $\lambda^j$ is prograde.
     
  Henceforth it will prove convenient to measure angles with respect
  to the line of sight.  In that case, $\angle V$ and $\angle V'$ are
  the phases of the TTV signals relative to the time when the
  longitude of conjunction points along the line of sight
  ($\lambda^j=0$).  In addition, the offsets $T$ and $T'$ then have
  the interpretation of being the time of any particular transit of
  the inner and outer planet, respectively.

In the  Appendix, we derive the  expressions for the complex TTV:
\begin{eqnarray} 
V&=& P{\mu' \over \pi
  j^{2/3}(j-1)^{1/3}\Delta} \left( -f-{3\over 2}{\E^*\over\Delta}
\right)\, ,
\label{eq:v}\\
V'&= & P'{\mu\over\pi j\Delta}
\left( -g+{3\over 2}{\E^*\over\Delta} \right)  \ ,\label{eq:vp}
\end{eqnarray} where $\mu$ is the mass ratio of the inner planet to the star
and $\mu'$ that of the outer planet, and $f$ and $g$ are  sums
of Laplace coefficients with order-unity values, as
listed in Table 
\ref{tab:simp}.
 Note that $f<0$ and $g>0$. 
When the pair is far from resonance, $|\Delta|$ is order unity, 
and the TTV expression reduces to $\sim P\mu$.
The key new dynamical  quantity that controls the TTV signal
is
 \be
\E\equiv fz_{\rm free}+gz_{\rm free}' \ ,
\label{eq:zfreeloc}
\ee 
 (with $\E^*$  its complex conjugate),
which is a linear combination of the free complex eccentricities
of the two planets. We proceed to define and describe the important
concept of free eccentricity.

\subsection{Free Eccentricity}
\label{subsec:free}

The complex eccentricity of a planet is
\be z = e e^{i \pomega}\, ,\label{eq:definez}\ee  where $\pomega$
  is the longitude of periapse. 
  Near a first-order
  mean-motion resonance, this can be decomposed into free and forced
parts,
\be z = z_{\rm free} + z_{\rm forced}\, . \label{eq:freeforce} \ee
The planet's forced eccentricity is forced 
by virtue of its companion's proximity to resonance.
    For the inner and outer
planets, 
   \beqn
\left(\begin{array}{c}z_{\rm forced} \\ z'_{\rm
      forced}\end{array}\right)=
-{1\over j\Delta}\left(\begin{array}{c}\mu'f(P/P')^{1/3} \\ 
\mu g
\end{array}\right)e^{i\lambda^j} \ ,
\label{eq:forced}
\eeqn (Eq. \ref{eq:zans}).  In the case  $\Delta > 0$,
 the forced eccentricity of the inner  planet is
aligned with the longitude of conjunctions, and that of the outer planet
is anti-aligned. The
situation is reversed for $\Delta < 0$.

\begin{figure}
\centerline{\includegraphics[width=0.45\textwidth,trim=0 0 0 0,clip]{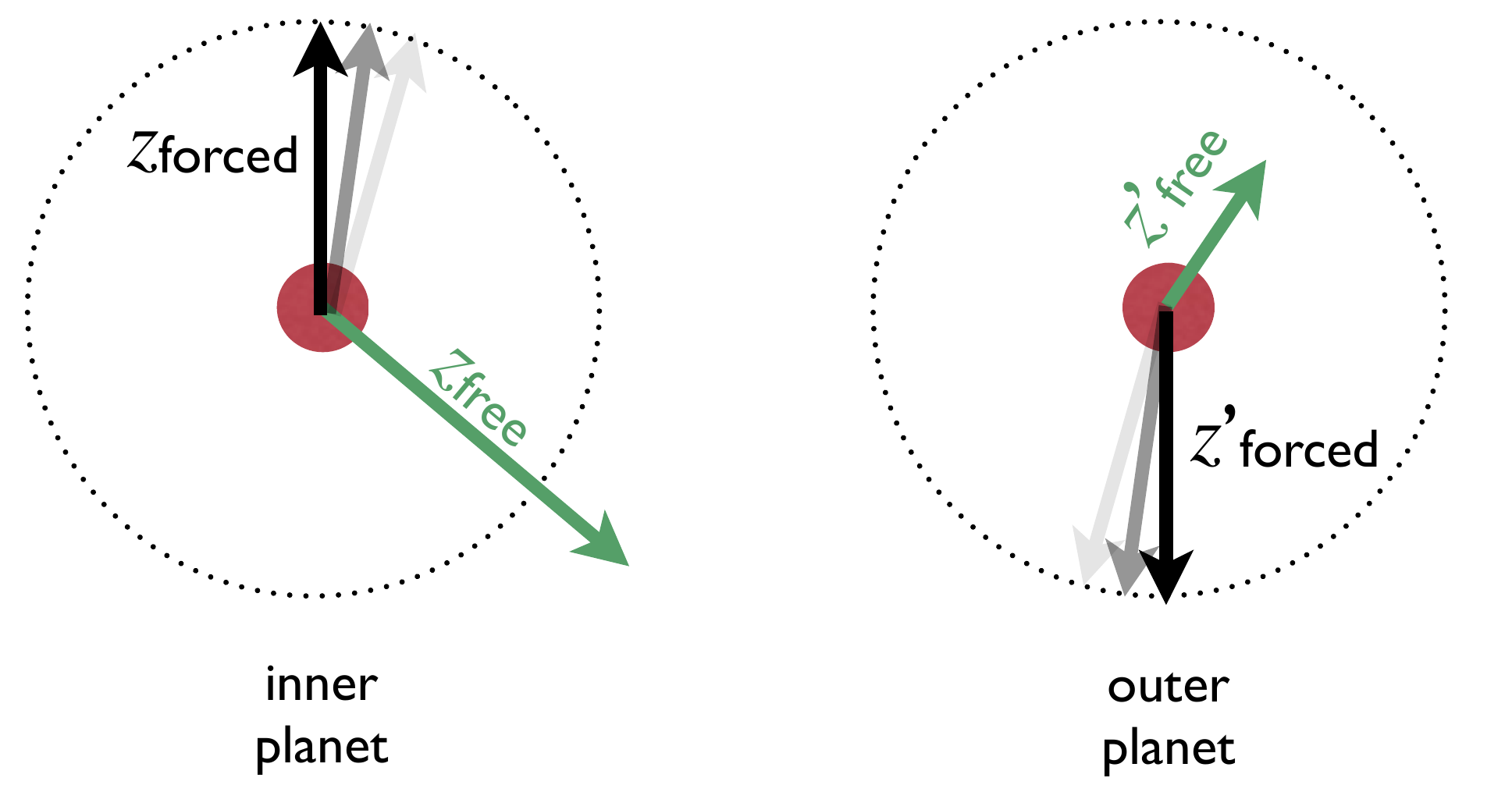}}
\caption{Schematic of free and forced complex eccentricities of a pair
  of near-resonant planets.  The complex eccentricity for each planet
  is a sum of free and forced components.  The free component is
  constant (on timescales $\ll$ secular times) with arbitrary
  amplitude and phase.  The forced component has magnitude determined
  by the planets' proximity to resonance, and phase that rotates in parallel with
  the longitude of conjunctions $\lambda^j$, with a super-period
  $P^j=P'/|j\Delta|$. }
\label{fig:ff}
\end{figure}

The free eccentricities can take arbitrary values.  They
represent the degrees of freedom associated with the non-circularity
of the orbits.  Although there are four such degrees of freedom
(corresponding to $e,\pomega,e',\pomega'$), they can only affect the
TTV signal through the linear combination of Equation
(\ref{eq:zfreeloc}). 
The complex free eccentricities precess on
  the secular timescale, $\sim P/\mu$, which is much longer than the
  super-period. So during any short time-span TTV observation, they
  can be taken as constant.  See Fig. \ref{fig:ff} for a cartoon
illustration of these concepts.

\subsection{Interpreting TTV}
\label{subsec:meaning}


Given two transiting planets, 
how much can be learnt 
about their parameters from measuring their TTV?
The analytical expressions allow for straightforward answers,
sparing one the abstruseness of N-body simulations.
If both transits are observed, then $P$, $P'$, $T$, $T'$ are known, 
even in the absence of observed TTV.
These yield $\Delta$ and $\lambda^j$. The TTV signals then allow one to
measure four quantities, the real and imaginary parts of
  $V$ and $V'$ (via Eqs. \ref{eq:tv}--\ref{eq:tvp}).
  Therefore the  four
unknown parameters ($\mu$, $\mu'$, and the real and imaginary parts of
$\E$) could  in principle be inferred by inverting
Equations \refnew{eq:v}-\refnew{eq:vp}.
 However, as we discuss in the following, degeneracies often arise that
prevent unique inversion.

 The {\it amplitudes} of the complex TTV 
  are
    \beqn
|V|\sim  P{\mu'\over|\Delta|}(1+{|\E|\over|\Delta|}) \ .
\label{eq:amp}
\\
|V'|\sim  P'{\mu\over|\Delta|}(1+{|\E|\over|\Delta|}) \ ,
\label{eq:ampp}
\eeqn after dropping order-unity coefficients; 
 we also assume for
  the purposes of the present discussion that $|\E/\Delta|$ is either
  very large or very small ($\gg 1$ or $\ll 1$).   We infer
  that the TTV amplitude of the inner planet yields an upper limit on
the mass of the outer planet, $\mu'\lesssim |\Delta| |V|/P$, and there
is a corresponding limit on the inner planet's mass.\footnote{  The
    TTV amplitude yields the mass if $|\E/\Delta|\ll 1$ and an upper
    limit on the mass if $|\E/\Delta|\gg 1$. But if $|\E/\Delta|\sim$
    unity, then the mass could greatly exceed that nominal upper limit
    provided 
     the complex phase of $\E$ is very nearly $0$ or
      $\pi$ (depending on the sign of $\Delta$). However, as we argue
    below, such a coincidence would be rare (Fig. \ref{fig:dists}).  
} But the value of $\mu'$ cannot be extracted from the TTV amplitude
without knowing the value of $|\E|$: a smaller $\mu'$ can be
compensated for by a higher $|\E|$ without affecting the amplitude of
$V$.  Similarly, if both planets have measured TTV's, one can
determine the ratio of their masses from the TTV amplitudes (within
order unity constants), but not the mass of either individually,
without knowing $|\E|$.

  Can the {\it phases}\footnote{Recall that we define the phases
    ($\angle V$ and $\angle V'$) to be relative to the time when the
    longitude of conjunctions points along the line of sight
    ($\lambda^j=0$). With this definition, the phases can be
    determined from observed TTV signals with no ambiguity about the
    origin of time or angle.  } be used to determine mass and free
  eccentricity uniquely?  This is in general impossible.  From
  Equations (\ref{eq:v})--(\ref{eq:vp}), the two planets' TTV signals
  are exactly out of phase with each other (anti-correlated) in either
  the limit that $|\E| \ll |\Delta|$ or $|\E| \gg |\Delta|$. This
  feature has been noted before and has been used to help confirm some
  Kepler planets \citep[][]{SteffenIII}.  In either limit, TTV signals
  only provide three independent quantities ($|V|$, $|V'|$ and one
  phase), and hence a degeneracy remains between planet mass and free
  eccentricity. To break it, additional information or assumptions are
  required, e.g., radial velocity measurements or the stability of the
  planetary system \citep[][]{CochranKepler18,FabryckyIV}. An
  alternative solution would be if TTV phases can be measured very
  accurately to discern the small deviation from anti-alignment.

 Nonetheless,  
  the phases  contain important information, and in certain circumstances they 
  {\it can}   help break
  the degeneracy between mass and eccentricity.
   For ease of discussion, we first define
   \beqn
   \phi_{\rm ttv}&\equiv& \angle (V\times {\rm sgn}\Delta) \label{eq:phi} \\
   \phi_{\rm ttv}'&\equiv&\angle (V'\times {\rm sgn}\Delta) \label{eq:phip} \ ,
   \eeqn
   where ${\rm sgn}\ \!\Delta=1$ if $\Delta>0$, and -1 otherwise.
   With these definitions, 
   $\phi_{\rm ttv}=0$ and $\phi_{\rm ttv}'=180^o$ when $\E=0$,
    independent of the sign of
   $\Delta$.
    Note that if $\phi_{\rm ttv}=0$, then
     $\delta t$ crosses zero from above to below
  whenever the longitude of conjunctions points along the line
  of sight (regardless of the sign
    of $\Delta$); similarly,  if $\phi_{\rm ttv}'=180^o$, 
    $\delta t'$ crosses zero from below when $\lambda^j=0$. 
    We consider the two possibilities:
  \bi
  \item 
   If the observed TTV's have a phase shift with
  respect to $\lambda^j$, that directly implies that  free
  eccentricities are present. A large phase shift implies
  $|\E|\gtrsim|\Delta|$.
\item If there is no phase shift (i.e., $\phi_{\rm ttv}=0$ and
  $\phi_{\rm ttv}'=180^o$), that does not necessarily imply that $|\E|
  = 0$.  Instead, $|\E|$ could be large but the phase of $\E$
  vanishes.  However, such a coincidence is unlikely.  Even if the
  phase of $\E$ vanished initially, secular precession would operate
  on the timescale $\sim P/\mu$ to randomize the phase. So if the free
  eccentricity is large ($|\E| \gg |\Delta|$), TTV phases should be
  randomly distributed between $0$ and $2\pi$. Conversely, if $|\E|
  \lesssim |\Delta|$, phase shifts should be small.  If many of the
  systems observed by Kepler have zero or near zero phase shifts, one
  could argue that most of them have small free eccentricities
  ($|\E|\lesssim |\Delta|$).  Such a result, if found true (see
  Section \ref{sec:appl}), has interesting implication for the
  dynamical history of these planets.  Moreover, for the task at hand,
  it would break the degeneracy in the mass determination: when
  $|\E/\Delta|$ is negligible in Eqs. \refnew{eq:v}-\refnew{eq:vp},
  $\mu$ and $\mu'$ are directly determined by the two TTV amplitudes.
  \ei

 Our above discussion is greatly aided by the analytical
 expressions. Using N-body simulations alone, it is difficult to
 elucidate the degeneracy between mass and free eccentricity.
 Moreover, the fact that TTV phases contain important information has
 hitherto been overlooked, only becoming transparent with the
 analytical formulae.

We conclude this subsection by comparing with previous work.  \cite{Agol}
estimate the TTV amplitudes for two near-resonant planets, assuming
that the planets' initial orbits are circular.  In that case, $z_{\rm
  free}=-z_{\rm forced}$ initially, and hence $\E\sim \mu/\Delta$
(Eq. \ref{eq:forced}).  Equations (\ref{eq:amp})--(\ref{eq:ampp}) then
imply $
|V| \sim P(\mu/|\Delta|)(1+\mu/|\Delta|^2) $, in agreement with
Equations (29)--(31) of \cite{Agol} in the two corresponding limits
($|\Delta|^2/\mu\ll 1$ and $\gg 1$).  We note, though, that there is
little reason why the initial orbits should be circular. If the
planets suffer weak damping (e.g., by tides or a disk), that would
tend to damp away the free eccentricities,  but the forced eccentricities would
  remain intact as long as proximity to resonance is maintained.
  \cite{Agol} also derive  analytically an expression for the TTV that is valid when
  $\E=0$
  (see their Appendix).  
\citet{nm08} and \cite{Nesvorny} derive a general analytic expression
for the TTV, but their expressions have not been applied to Kepler
planets due to their algebraic complexity, which arises partly because
they do not distinguish free from forced eccentricities.

\subsection{Testing with N-body Simulations}
\label{sec:nbody}

\begin{figure}
\centerline{\includegraphics[width=0.49\textwidth,trim=10 140 20 105,clip]{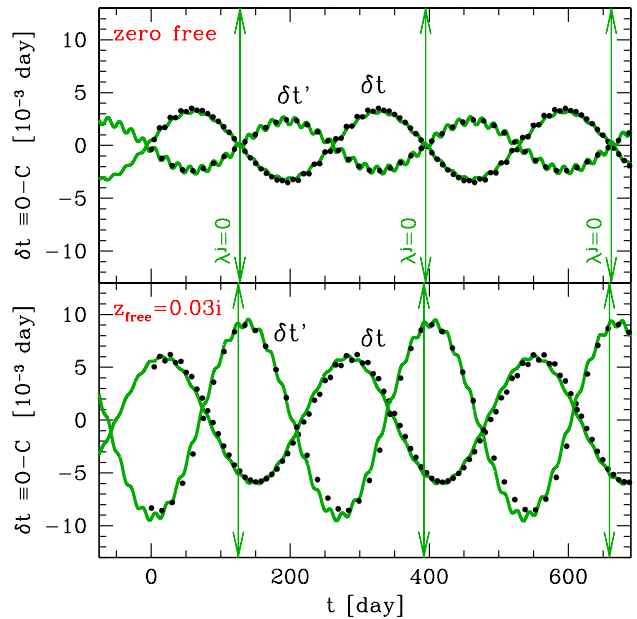}}
\caption{TTV's of a near-resonant planet pair, without (above) and
  with (below) free eccentricity.  Planet periods and masses are
  chosen to be close to those of Kepler-18c/d (see Table
  \ref{tab:param}) with $\Delta = -0.028$ relative to the 2:1
  resonance.
  The black points show the N-body simulated TTV signals for the two
  planets and the green curves the analytic expressions
  (Eqs. \ref{eq:tv}--\ref{eq:zfreeloc}).  The vertical arrows show the
  times at which the longitude of conjunctions points at the observer
  ($\lambda^j=0$). The duration between two such arrows is the
  super-period. In the top panel, the pair have zero free eccentricities, so their TTV
  cross through zero when $\lambda^j=0$, with the inner planet
  crossing from above, and the outer from below. The bottom panel
  shows the case when the inner planet has free eccentricity
  $z_{\rm free} = 0.03 i$, which  is  $\sim 20$ times
    greater in amplitude than its forced eccentricity and is of order $|\Delta|$.  In
  this case, the TTV's are no longer in phase with $\lambda^j$, 
    though the inner and outer TTVs are still anti-correlated. 
    The small rapid wiggles in the green curves are due to our inclusion
    of the 3:2 forcing term in addition to the 2:1;  this clearly has little 
    effect.}
\label{fig:nbody1}
\end{figure}

We test our TTV expressions with N-body simulations, choosing cases
that illustrate our discussion above. Figure \ref{fig:nbody1} shows
transit time variations from two separate N-body simulations for a
pair of planets that resemble Kepler-18 c/d, showing agreement with
the predictions of Equations (\ref{eq:tv})--(\ref{eq:zfreeloc}).
   In the top panel the planets have zero free eccentricity
and hence  TTV are in phase with $\lambda^j$, while in the bottom
panel the inner planet has a free eccentricity and hence the signals
are not in phase.  
Numerically, we reach the state of zero free eccentricity by first weakly damping
the planets' velocities to the local circular speed over a few million
orbits.  Many other forms of weak damping will also remove free
eccentricities. 

\begin{figure}
\centerline{\includegraphics[width=0.49\textwidth,trim=10 140 20 105,clip]{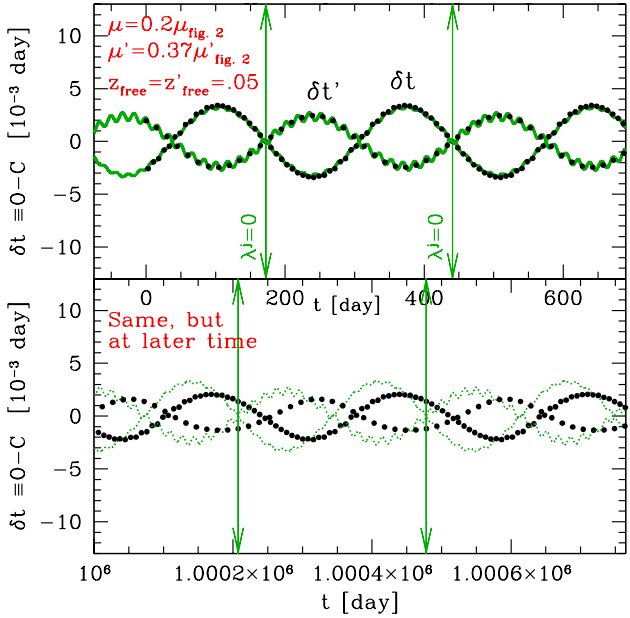}}
\caption{Degeneracies in extracting planet parameters using TTV
  signals. Here points result from N-body simulations, and green
  curves are theoretical results.  The top panel shows a pair of
  planets exhibiting the same TTV signals as those in the top panel
  of Fig. \ref{fig:nbody1}, but the ones here have reduced planet masses and
  non-zero free eccentricities (as marked).  The TTV's remain in phase
  with $\lambda^j$ because we deliberately chose free eccentricities with
  $\angle z_{\rm free}=\angle z'_{\rm free}= 0$ initially.  This panel
  illustrates that TTV measurements alone do not allow  unique inference of
  mass and free eccentricity. Additional constraints must  be imposed.  The points in the bottom panel show
  the same system but $10^6$ days later, when the phases of the free
  eccentricities have secularly precessed. The TTV's are no longer in
  phase with $\lambda^j$. The faint green curves in this panel are what the
  TTVs would look like for the system with
  zero free eccentricity
   (i.e, the system in the top panel of Fig. \ref{fig:nbody1}) --
   its TTV remains in phase.  Therefore the
  phase of the TTV provides a statistical way to break the
  degeneracies inherent in the TTV's.  }
\label{fig:nbody2}
\end{figure}

Figure \ref{fig:nbody2} illustrates the degeneracies inherent in
extracting planet parameters from TTV, and how these can be partially
removed with the phase information, as discussed in Section
\ref{subsec:meaning}.  In the top panel, we perform a simulation
similar to the one in the top panel of Figure \ref{fig:nbody1}, with
the same periods for the two planets, but in this case the masses of
the two planets are reduced by factors of $0.2$ and $0.37$.
Furthermore, after the free eccentricities are damped away the two
planets' complex eccentricities are increased by $z_{\rm free}=z'_{\rm
  free}=0.05$.  The resulting TTV's are almost identical to those in
the top panel of Figure \ref{fig:nbody1}, despite the vastly different
parameters.  Hence if one observed such TTV's, one could not determine
the planets' masses.  Nonetheless, the fact that the TTV's are in
phase with $\lambda^j$ in the top panel of Figure \ref{fig:nbody2} is
due to our judicious choice of $\angle z_{\rm free}=\angle z'_{\rm
  free}=0$.  Even though they are initially in phase, this cannot
remain true for long: the black points in the bottom panel show the
TTV's in the same simulation, but $10^6$ days later, by which time the
two planets' free longitudes of periapse have precessed by $\sim
70^o$.  As a result, the TTV's are no longer in phase with
$\lambda^j$.  By contrast, the TTV's of pairs with zero free
eccentricity would always remain in phase with the longitude of
conjunction (faint green line in bottom panel of
Fig. \ref{fig:nbody2}).  This example shows that if 
 many systems are
 observed to have 
  small
 TTV phase
 shifts, 
 one could argue that 
most
    planetary systems likely have small free eccentricities, 
  although one cannot be certain
    for any
  particular system.
   Furthermore, if the free eccentricities
  are small, one could
   determine planet masses from TTV
data 
 (see Fig. \ref{fig:dists}). 

\section{Applying to Kepler pairs: mass and free eccentricity}
\label{sec:appl}
\label{sec:data}

Equations (\ref{eq:v})--(\ref{eq:vp})  show clearly how planet mass and
eccentricity affect the TTV signals.
The task to invert observed TTV signals to obtain physical parameters
is now almost trivial.

Transit times have been published for $ 13$ Kepler systems with confirmed
planets
\citep[e.g.,][]{CochranKepler18,Fabryckyetal12,Ford12}.  Here, we
apply our formula to those near 2:1 or 3:2 resonances.
We further restrict to systems that have gone through at least one
complete TTV cycle, leaving us with 6 systems, all of which have inner
period $<10$ days.  

\begin{figure}
\centerline{\includegraphics[width=0.49\textwidth,trim=10 300 50 80,clip]{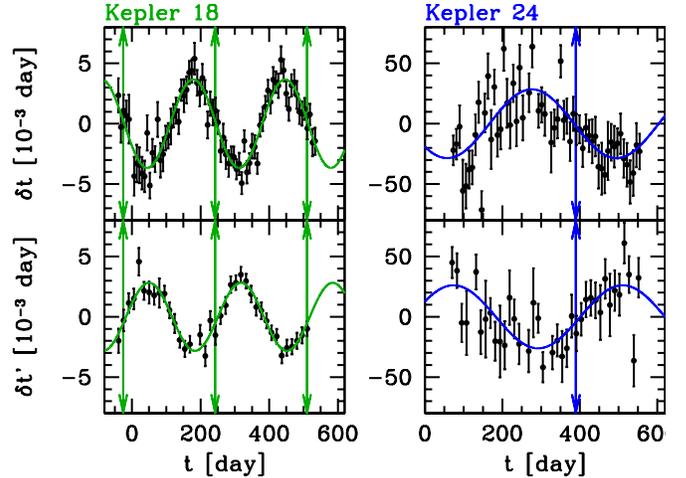}}
\caption{TTV data (black points with errorbars), best-fit theoretical
  curve, and times when the longitude of conjunction points at the
  observer, i.e., $\lambda^j=0$ (vertical arrows).  These two systems
  have near-zero phase shift, as evidenced by the fact that the inner
  planet's TTV (upper panel) crosses through zero from above at times
  when $\lambda^j=0$; and the outer planet's TTV crosses through zero
  from below.  Hence these systems likely have zero free eccentricity.
  Kepler 18 data from \cite{CochranKepler18}, and Kepler 24 data from
  \cite{Ford12}. The former system lies inside the 2:1
    resonance, while the latter one lies  outside the 3:2.
      }
\label{fig:figttva}
\end{figure}

\begin{figure}
\centerline{\includegraphics[width=0.49\textwidth,trim=10 300 50 80,clip]{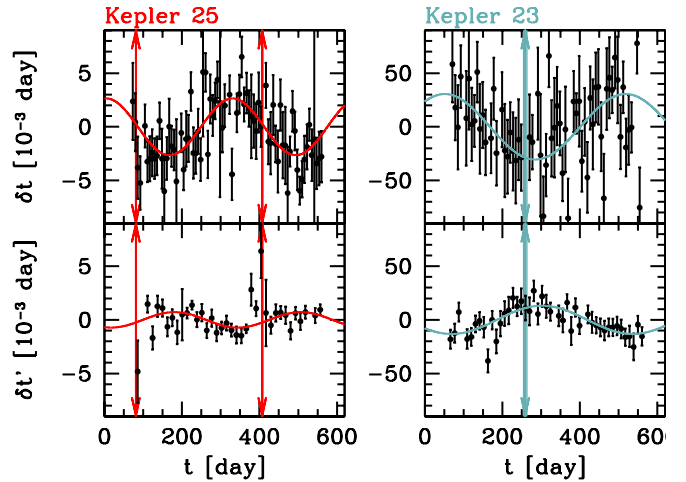}}
\caption{TTV data and fits.  Similar to Figure \ref{fig:figttva} but
  for Kepler 25 \citep[data from][]{SteffenIII}) and Kepler-23
  \citep[data from][]{Ford12}. Kepler 25 has zero phase shift, but
  Kepler 23 has a non-zero phase shift.}
\label{fig:figttvb}
\end{figure}

\begin{figure}
\centerline{\includegraphics[width=0.49\textwidth]{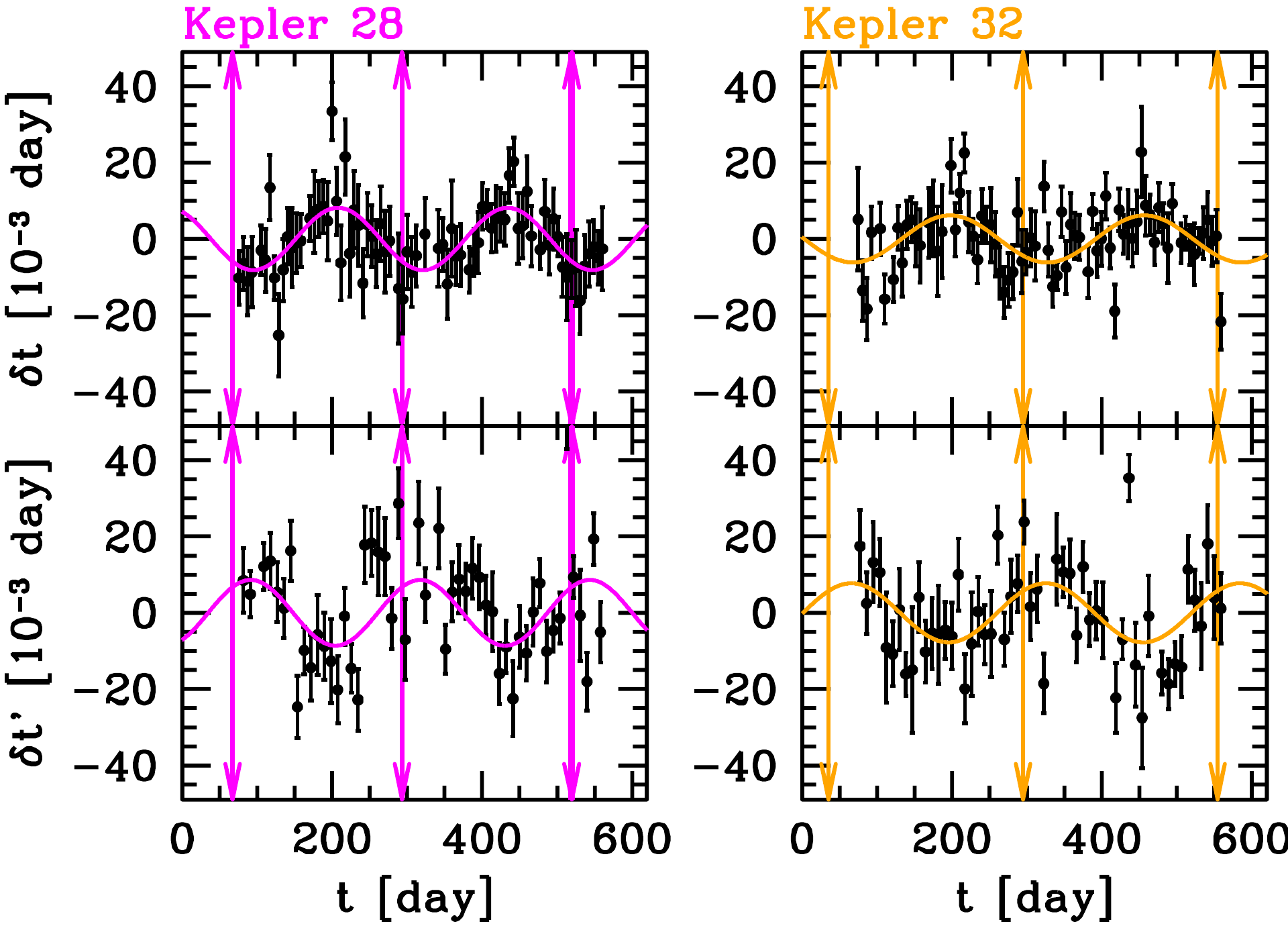}}
\caption{TTV data and fits for Kepler-28 \citep[data
  from][]{SteffenIII} and Kepler-32 \citep[data from][]{FabryckyIV}.
  Both these systems have non-zero phase shifts.  }
\label{fig:figttvc}
\end{figure}

Our results are depicted in Figures
\ref{fig:figttva}--\ref{fig:figmassrad} and Tables
\ref{tab:param}--\ref{tab:masses}.  All six short period systems
appear consistent with having small free eccentricity $|\E| \lesssim
0.01$, with three of them consistent with zero free eccentricity.
Before discussing these results, we describe our method.


\begin{table*}
\begin{center}
\begin{minipage}{140mm}
\caption{Complex TTV for Six Kepler Systems}
\begin{tabular}{|c|cccc|cccc|cc|}
\hline
Kepler ID &         j:j-1&   $P$ [d]             &               $P'$   [d]       &                $\Delta$         &         $|V|$  [d]
 &            $\phi_{\rm ttv}$              &            $|V'|$   [d]           &                $\phi_{\rm ttv}'$
  & $\chi^2_{\rm dof}$ & $\chi'^2_{\rm dof}$
  \\ \hline
     18c/d& 2:1    &  7.642 & 14.86 &-0.028   & $0.0037(\pm7\%)$  &$-4.3^o\pm 4^o$   & $0.0028(\pm 10\%)$ & $169^o\pm 5^o$ & 0.95 & 0.71
 \\
 24b/c& 3:2 & 8.146  & 12.33 & 0.0094 & $0.028(\pm 20\%)$ & $-3.9^o\pm 12^o $ 
&$0.026(\pm 20\%)$ &  $180^o\pm 16^o$ & 1.95 & 1.28
\\
     25b/c & 2:1&6.239&12.72&0.0195 &$0.0026(\pm 20\%)$&$5.7^o\pm 12^o$&
$0.00072(\pm 30\%)$&$200^o\pm 22^o$ & 1.19 & 1.60
\\
  & & & & & & & &  & & \\
     23b/c &3:2& 7.107&10.74&0.0077&$0.031(\pm 40\%)$&-$68^o\pm 18^o$&$0.013(\pm 30\%)$&$120^o\pm 17^o$ & 1.88 & 0.93
    \\
     28b/c&3:2&5.912&8.986&0.013&$0.0082(\pm 20\%)$&$-50^o\pm 14^o$&$0.0086(\pm 30\%)$&$130^o\pm 16^o$   & 0.87 & 3.43
    \\
     32b/c&3:2&5.901&8.752&-0.011&$0.0062(\pm 30\%)$&$45^o\pm 15^o$&$0.0077(\pm 30\%)$&$228^o\pm 19^o$ &  1.33 & 2.60
   \\
       \hline
\end{tabular}
\tablecomments{  Columns labelled $|V|$, $\phi_{\rm ttv}$,
$|V'|$, and $\phi_{\rm ttv}'$ are the amplitudes and phases of
  the complex TTV, extracted from the transit times with a least
  squares fit (see text); the phases are defined in Equations
  (\ref{eq:phi})--(\ref{eq:phip}). 
   The last two columns are the  values of $\chi^2/$(deg. of freedom) from the fits for the inner and outer
  planet.
   The phases for the first three
  systems  in the table are consistent with zero free eccentricity ($\phi_{\rm
    ttv}=0$, $\phi_{\rm ttv}'=180^o$), and those of the latter three
  are not.  Errors on the TTV amplitudes and phases are at 68\%
  confidence.  }
 \label{tab:param}
\end{minipage}
\end{center}
\end{table*}

\begin{table}
\begin{center}
\begin{minipage}{80mm}
\caption{Planet Masses}
\begin{tabular}{|c|cc|cc|}
\hline
 & \multicolumn{2}{c|}{analytical Mass [$M_\oplus$]} & \multicolumn{2}{c|}{N-body Mass [$M_\oplus$]} \\
Kepler ID &       $m_{\rm nominal}$ & $m'_{\rm nominal}$ &  $m$ & $m'$ 
 \\ \hline
     18c/d& $20.2\pm 1.9$&$17.4\pm 1.2$ 
& $17.3 \pm 1.7$ & $15.8 \pm 1.3$
 \\
 24b/c& $28.4\pm 5.9$&$50.4\pm 7.9$
 & $56.1 \pm 15.8$ & $102.8\pm 21.4$
\\
     25b/c &$7.13\pm 2.5$&$13.1\pm 2.6$ 
& $8.1 \pm 3.1$ & $13.3 \pm 3.9$
\\
  & & & & \\
     23b/c &$14.7\pm 3.8$&$55\pm 22$ 
& $4.8 \pm 15.6$ & $15.0 \pm 49.8$
\\
     28b/c& $14.8\pm 4.2$&$22.9\pm 5.6$ 
& $3.8\pm 6.9$ & $4.9 \pm 9.3$
\\
     32b/c&$6.0\pm 1.9$&$7.59\pm 2.0$ 
& $7.2 \pm 4.1$ & $5.2 \pm 3.5$
\\
       \hline
\end{tabular}
\tablecomments{ The nominal masses $m_{\rm nominal}$ and $m'_{\rm
    nominal}$ are calculated using Equations
  (\ref{eq:mv})--(\ref{eq:mvp}); these are likely modest overestimates of the true masses.  For the stellar masses, and the N-body
  determined planet masses,
  we use those listed in Table 5 of \cite{FabryckyIV}, except for
  Kepler-18 for which we use values from \cite{CochranKepler18}.
   }
 \label{tab:masses}
\end{minipage}
\end{center}
\end{table}

\subsection{Extracting the Complex TTV }
\label{sec:method}

Given a sequence of transit times for two planets in a system, we
obtain the parameters $P,P',T,T',V,V'$ in Equations
(\ref{eq:tv})--(\ref{eq:lams}) as follows.  We write the transit times
for the inner planet as
 (Eq. \ref{eq:tv})
 \be t_{\rm trans}=T+Pi_{\rm trans}+{\rm
  Real}(V)\sin\lambda^j+{\rm Imag}(V)\cos\lambda^j
\label{eq:ttrans}
\ee
where $i_{\rm trans}=0,1,\cdots$ is the transit number.
Since $|V|\ll P$, we proceed in two steps.
First, we fit the inner planet's $t_{\rm trans}$ vs. $i_{\rm trans}$ with a straight line, 
thereby extracting $P$ and  $T$, and also do the same for the outer planet, extracting $P',T'$.
Our fits are done by  linear least squares \citep[e.g.][]{press}.
Second,  we fit 
for the four parameters explicit in Equation (\ref{eq:ttrans}) (i.e.,
$T, P, {\rm Real}(V), {\rm Imag}(V)$) using a second least squares fit
for the inner planet's transit times; in this fit, $\lambda^j$ is
determined by Equations (\ref{eq:lamj})--(\ref{eq:lams}) at times
$t_{\rm trans}$, where the parameters $T,P,T',P'$ are the ones
obtained from the first fits.  This refitting for $P$ 
helps to remove
 the small linear trend that remains after the first fit.  We repeat
for the outer planet.  This procedure implicitly assumes that the
period of the TTV curve is  the super-period
(Eq. \ref{eq:psuper}).  That this assumption is valid for the six
Kepler pairs of interest can be seen by comparing the fit to the data
(Figures \ref{fig:figttva}--\ref{fig:figttvc}).  The results of the
fits
 are listed in Table \ref{tab:param}.

\begin{figure}
\centerline{\includegraphics[width=0.49\textwidth,trim=10 140 20 100,clip]{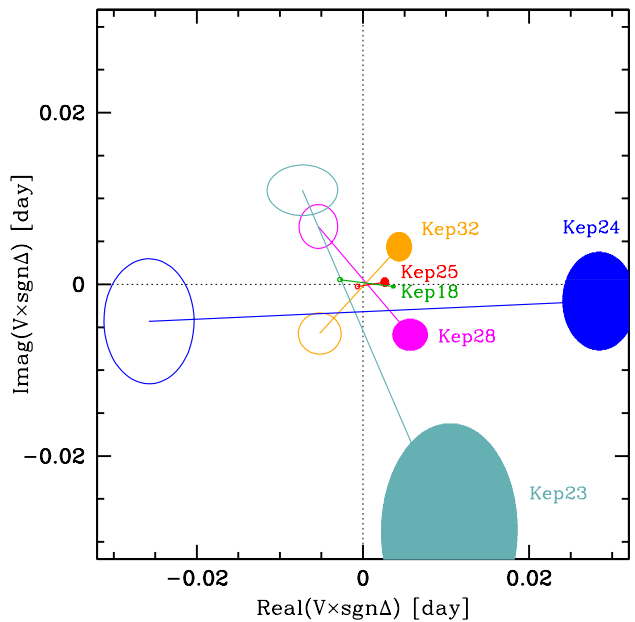}}
\caption{Best-fit complex TTVs, plotted in the complex plane.  Here,
  ${\rm sgn} \Delta$ stands for the sign of $\Delta$ -- this term is
  included so that pairs with zero free eccentricity would lie on the
  horizontal dotted line with the inner planet to the right,
  regardless of the sign of $\Delta$.  The filled ellipses are the
  $68\%$ 
  confidence regions for the inner planet of each pair ($V$), and the
  open ellipses are for the outer planet ($V'$), connected here as
   dumbbells.  Kepler 18, 24, and 25 are all consistent with
  having zero free eccentricities, while three other systems shown
  here are discrepant (Kepler 23, 28, 32).  But even for these, the
  inner planets still lie mostly to the right suggesting that the free
  eccentricity is small ($|\E| \lesssim |\Delta|$).   }
\label{fig:figell} 
 \end{figure}

\begin{figure}
\centerline{\includegraphics[width=0.49\textwidth,trim=10 140 20 100,clip]{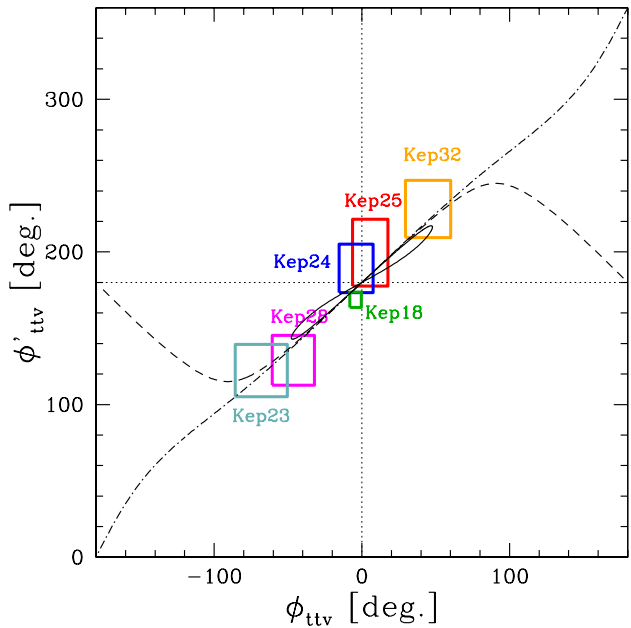}}
\caption{ Phases of the complex TTV. The x-axis is the phase of
    the inner planet's TTV, and the y-axis is for the outer planet
    (Eqs. \ref{eq:phi}--{\ref{eq:phip}}).
 A system with zero free
  eccentricity would lie at the center of the plot.
The width and height of each rectangle denote the $68\%$ confidence
limits measured from Kepler data for the $6$ systems in this
study. The three curves represent the theoretical prediction for the
TTV phases when $|\E| = \Delta$ (solid curve), $|\E| = 1.5 \Delta$
(dashed curve) and $|\E| = 2 \Delta$ (dot-dashed curve) for systems
near the 3:2 resonance. It appears that all systems can be explained
with small free eccentricity, $|\E| \leq |\Delta|$.  }
\label{fig:ph}
\end{figure}

To obtain error estimates for the best-fit values, we use the
covariance matrix provided by the second least-square fit and assume
that the quoted 1-$\sigma$ errors in the  transit time data are
independent and Gaussian. The errors in periods ($P,P'$) and fiducial
transit times ($T,T'$) are negligibly small (fractional error
    $\ll 10^{-4}$).   All of our errors are quoted at 68\% confidence.

 Figure \ref{fig:figell} depicts the inferred values of $V$ and $V'$ for the twelve
planets.   
 Figure \ref{fig:ph}, which plots just the phases,
 summarizes the main result of our analysis:   three of the
systems
are consistent with having zero phase and three  are not. But
all six systems
have  $|\phi_{\rm ttv}|\lesssim 90^o$. 
This strongly suggests that $|\E|\lesssim|\Delta|\sim 0.01$; otherwise, 
 the six systems would have random phases
between $-180^o$ and
$180^o$.

 Table \ref{tab:masses}  lists the masses
that would be inferred by setting $\E=0$ in Equations
(\ref{eq:v})--(\ref{eq:vp}), i.e., \beqn m_{\rm nominal}&\equiv& M_*
\left|{V'\Delta\over P'g}\right|\pi j \label{eq:mv}
\\
m'_{\rm nominal}&\equiv& M_* \left|{V\Delta\over Pf}\right|\pi
j^{2/3}(j-1)^{1/3} \ . \label{eq:mvp} \eeqn 
  Figure \ref{fig:figmassrad} displays these
nominal masses  versus planet radii. 
The true mass is related to the nominal mass by
\be
{m\over m_{\rm nominal}}={1\over \left|1-\E/(2g\Delta/3)  \right| } \ ,
\label{eq:mmnom}
\ee
and similarly for the outer planet after replacing $g\rightarrow -f$.
 Although $\E$ is not known, the inference that
  $|\E|\lesssim|\Delta|$ implies that the nominal masses are
typically close to, but a little bigger than, the planets' true masses
(Section \ref{sec:measure}).

\subsection{Systems with negligible TTV phase shift: Kepler-18, 24, 25}
\label{subsec:zerofree}

From Figures \ref{fig:figttva}--\ref{fig:figttvb}, we see that Kepler
18, 24, and 25 all have very small phase shifts: $\delta t$ crosses
through zero from above at the times when $\lambda^j=0$, and $\delta
t'$ crosses from below. 
 This can also be seen in Figures
\ref{fig:figell}--\ref{fig:ph}.

{\bf Kepler 18: } This system has currently one of the best measured
TTV, and its TTV phase lies very close to zero. If we {\it assume}
that this system has zero free eccentricity, then we may deduce the
masses of the two planets with small error bars.  These values lie
close to those obtained using an N-body fit (Table \ref{tab:masses}).

  But to obtain these mass estimates and small error bars, we must
  assume that the free eccentricities vanish (or $|\E|\ll |\Delta|$).
  Without this assumption, the masses of the planets are degenerate
  with the free eccentricity:
  one could choose an infinite sequence of $\E$ with $\angle \E
  \approx 0$ or $\pi$ to reproduce Kepler 18's TTV signals,  each
  corresponding to a different set of masses.
  However, the secular precession of $\E$ implies that a specially
  aligned $\E$ would be unlikely. It is far more likely the free
  eccentricities are very small, $|\E| \lesssim |\Delta| \sim 0.03$.

  Without explicitly assuming that the free eccentricity is small,
  \cite{CochranKepler18} obtain mass estimates and errors similar to
  ours by fitting the TTV signal (alone) with N-body simulations
  (Table \ref{tab:masses}).  We suggest this result is fortuitous:
  their N-body simulations have not exhaustively searched all possible
  mass-eccentricity combinations.   Nearly the same TTV signals can be produced
  by much lower planet masses, as can be seen by comparing the top panels
  of Figures \ref{fig:nbody1} and \ref{fig:nbody2}.  
  In truth, the TTV's in those
  two panels differ slightly. 
  Such  ``chopping'' \citep{HolmanKepler9,FabryckyIV}
  or other non-sinusoidal behavior
   might be used to break
  the degeneracy between mass and eccentricity.  But it is not clear if  the data
  is of high enough quality to distinguish
   the difference
    between those two panels.
 By contrast,  in Kepler 36 for example,  which lies close to a high $j$ resonance (7:6),  
  the TTV signals are significantly non-sinusoidal, and that likely allowed the 
  N-body fit to break the degeneracy
   \citep{carterkepler36}.

      In the case of Kepler 18, additional  non-TTV information,
      such as radial velocity measurements and theoretical
      expectations for planet density allow one to eliminate most of
      the solutions. But  that would not be
      possible for most systems.  This example accentuates the
      importance of an analytical understanding:  the degeneracy
        between mass and eccentricity are explicit in the formula.

  {\bf Kepler 25:} The TTV phases of Kepler 25b/c are also consistent
  with zero, a strong indication that the planets' free eccentricities
  are very small. From our fits for $|V|$ and $|V'|$ listed in Table
  \ref{tab:param}, we derive nominal masses for the 
  planets as listed in Table \ref{tab:masses}.  These are likely
   close to
   the
  true planet masses.  They are consistent with those from N-body fits
  (Table \ref{tab:masses}).
  However, as for the Kepler-18 case, the N-body fit without any
  assumption on the value of the free eccentricity is incomplete. An
  alternative solution that matches the observations is, e.g., inner
  and outer masses of $0.59 M_\oplus$, $3.0 M_\oplus$ and $\E =
  -0.05$.

{\bf Kepler 24:} Similar arguments apply to Kepler 24, for which 
the phases are also consistent with zero.  We derive masses of
$28.4\pm 5.9M_\oplus$ and $50.4\pm 7.9M_\oplus$ for the inner and
outer planets, when we assume that the free eccentricity vanishes.
\citet{FabryckyIV} list masses from N-body fits of $56.1 \pm 15.8
M_\oplus$ and $102.8 \pm 21.4 M_\oplus$, and large eccentricities, of
order $0.4$ and $0.3$ for the inner and outer planets.  However, the
TTV signal   suggests smaller
  eccentricities.\footnote{The
  solution of \cite{FabryckyIV} for Kepler 24 yields $|\E|\ll e$:
  their eccentricity vectors nearly cancel in the combination
  $fz+gz'$.  That is likely why their masses are only twice as large
  as ours rather than $\sim e/|\Delta|\sim 20$ times larger.  We note
  also that  the TTV super-period 
   from the periods in Table \ref{tab:param} ($P^j\sim$450 d)
   agrees with that of \cite{Ford12}, but
   disagrees with  that from Table 6 in \cite{FabryckyIV} by
  60\% because the latter uses osculating periods (D. Fabrycky, personal communication).  \citet{Ford12} report
  two more candidates in the system, KOI-1102.04 and KOI-1102.03. If
  real \citep[they are absent from the table in][]{Batalhaetal12},
  they are near resonant with the two planets here but with $\Delta
  \sim 3\%$ or $4\%$, respectively. These should not significantly
  affect the TTV amplitude.}

\begin{figure}
\centerline{\includegraphics[width=0.49\textwidth,trim=10 140 20 100,clip]{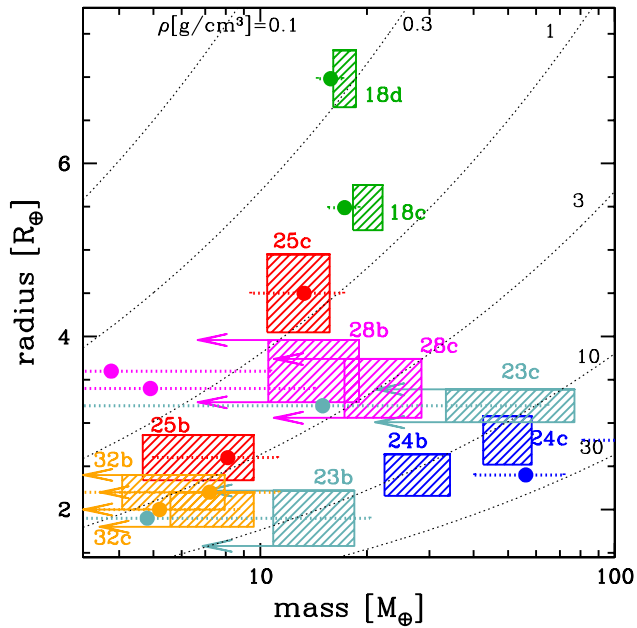}}
\caption{ Nominal mass
    vs. radius  for  planets in
  the 6 Kepler systems we examine: The shaded rectangles show our
  values for the nominal masses (Eqs. \ref{eq:mv}--\ref{eq:mvp}),
  plotted against planet radii, encompassing our error estimates.  The
  nominal masses are likely  modest overestimates of the true
  masses, by a factor of 1--2 (Fig. \ref{fig:dists}).
     For comparison, masses determined in the
  literature with N-body simulations are marked with filled circles.
  For radii, see references in
  Figs. \ref{fig:figttva}-\ref{fig:figttvc}.  The error estimates on
  radii are unpublished for Kepler 24, 25, 23, and 28, and hence
  assumed to be 10\% for this plot.  }
\label{fig:figmassrad}
\end{figure}

With planet radii of $2.4$ and $2.8 R_\oplus$
\citep{Ford12,Batalhaetal12}, our  nominal masses imply
  surprisingly high densities of $
  \sim 12 \g/\cm^3$, comparable to the density of lead
  (Fig. \ref{fig:figmassrad}).  But the true masses and densities
  could be smaller than these nominal values by a factor of $\sim 2$
  if the free eccentricities were $\sim 0.01$, even though
  in that case it would have to be a coincidence that the phase nearly
  vanishes.

\subsection{Systems with TTV phase shift: Kepler-23,28,32}
\label{subsec:somefree}

Three out of six systems we analyze are inconsistent with zero free
eccentricity: their TTV phases differ significantly from $\phi_{\rm
  ttv}=0$ and $\phi_{\rm ttv}'=180^o$.  
All are near 3:2 resonances. Planet masses in these cases cannot be
reliably determined.
However, as is apparent from Fig. \ref{fig:ph}, there is an absence of
systems  with large
  phase shifts.

    {\bf Kepler 23:} This pair has the largest phase shift, $\sim
    70\deg$.  The nominal masses correspond to high planet densities
    $\sim 10\g/\cm^3$, comparable to silver.   Equation
      (\ref{eq:mmnom}) shows that if one takes $|\E| = 4 \Delta =
    0.03$,  for instance, then the real masses would be three
    times lower than the nominal ones.

    {\bf Kepler 28 and 32:} The nominal masses in these two cases
     yield planet densities $\sim 3\g/\cm^3$. The true masses
    are likely smaller  (Eq.  \ref{eq:mmnom}).  However, if the
    collective behavior of these TTV pairs indeed implies that $|\E|
    \approx |\Delta|$, then the true masses lie close to the nominal
    ones.

\begin{figure}
\centerline{\includegraphics[width=0.49\textwidth,trim=10 140 20 100,clip]{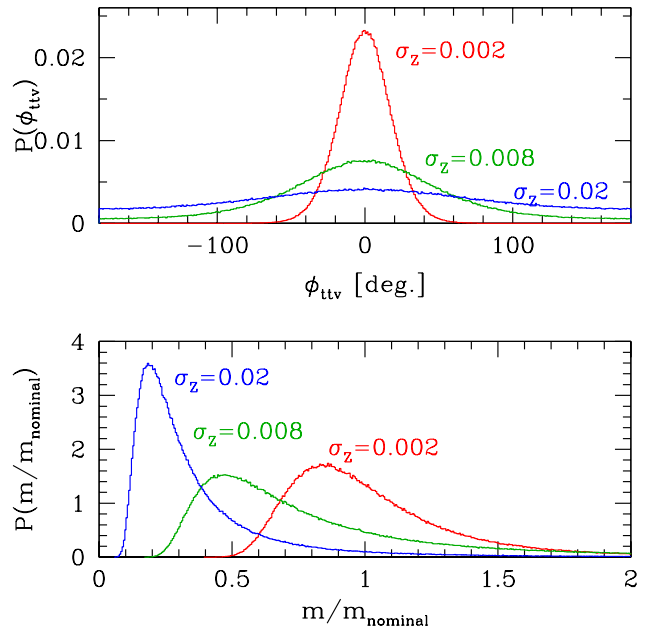}}
\caption{ Probability distributions of TTV phase and true planet
  mass, assuming that the real and imaginary parts of $\E$ are
  independent Gaussians, with r.m.s.  $\sigma_Z$.  The top panel shows
  the resulting distributions of TTV phases (via Eq. \ref{eq:v}) for
  three values of $\sigma_Z$, after taking $|\Delta f 2/3|=0.008$;
  equivalently, the three distributions have $\sigma_Z/|\Delta f
  2/3|=0.3$, 1, and 3.  The bottom panel shows the resulting
  distributions of $m/m_{\rm nominal}$ (via Eq. \ref{eq:mmnom}).  }
\label{fig:dists}
\end{figure}

\subsection{Extracting Mass and Eccentricity}
\label{sec:measure}
\label{sec:extract}
Given the measured TTV phases and nominal masses 
(Figs. \ref{fig:ph} and \ref{fig:figmassrad}), what can be concluded about the planets' 
eccentricities and  masses? As emphasized above, the 
mass-eccentricity
degeneracy can only be broken in 
a statistical way. 
Because of the small number of systems analyzed in this paper, we
leave a proper statistical study to future work.  Instead, here we
model our results by assuming that the free eccentricities of all
planets   are randomly drawn from a Rayleigh distribution
\citep[][]{bt2}
with random phases, i.e., that the real and imaginary parts of the
complex free eccentricities ($z_{\rm free}\equiv e_x+ie_y$) are drawn from independent Gaussians
\begin{equation}
P(e_x) = {1\over{\sqrt{2 \pi}}\sigma} \exp(- {{e_x^2}\over{2\sigma^2}})\, ,
\label{eq:psigma}
\end{equation}
and $e_y$ similarly.
The top panel of Figure \ref{fig:dists} shows the resulting phase
distribution for three values
 $\sigma_Z = \sigma \sqrt{f^2 + g^2}$
(assuming $f\Delta$ is fixed).  From the fact that half
of the systems have phases $|\phi_{\rm ttv}|\lesssim 10^o$, we infer
that $\sigma_Z\sim 0.002-0.008$.  Therefore the free eccentricities of
the planets are quite small, $\lesssim 0.006$.
  The
bottom panel of Figure \ref{fig:dists} shows the distribution of the
ratio of true to nominal mass (Eq. \ref{eq:mmnom}).  For the inferred
$\sigma_Z$, we see that the true masses are comparable to the nominal
masses, within around a factor of 2.  We emphasize, however, that it
is not necessarily the case that the free eccentricities in these six
systems are drawn from the same distribution. Instead, it could be
that the three systems consistent with zero phase all have precisely
$\E=0$. In that case, the nominal masses for those systems would be
their true masses.  A better assessment of probable values awaits
future data analysis.  

\section{Summary}
\label{sec:sum}

\subsection{Analytical TTV}

We have derived simple analytical expressions for the TTV from two
planets near a first order mean motion resonance
(Eqs. \ref{eq:tv}--\ref{eq:zfreeloc}).  These show that the amplitude
and phase of the TTV depend on both planet mass and free eccentricity.
There is an inherent degeneracy between mass and free eccentricity
which in general prevents either from being determined independently
of the other. This degeneracy, however, may be (partially) broken
under certain circumstances, based on probability arguments.

  There is a special moment in time when the longitude of conjunction
  points along the line of sight. If the phase of TTV is zero relative
  to this time, then it is  likely that the free eccentricity in
  the system is zero. 
  Moreover, if the free eccentricity is zero, the
    TTV amplitudes can  be used to uniquely
  determine planet masses.

  When applying this technique to six published systems, we find that
  three of them are consistent with zero TTV phase, while the other
  three deviate by less than a radian. This clustering around zero
  phase can be most naturally explained if all systems have free
  eccentricity of order a percent or less---which is comparable to the
  typical distance to resonance in these systems.  Furthermore, because the free
    eccentricities are small, the nominal masses determined by TTV are
    likely close to the true masses, within a factor $\lesssim 2$.

  Without the analytical TTV expressions and relying only on N-body
  simulations, it would be hard to reach these conclusions.

\subsection{Implications of Small Free Eccentricities}

The  very small free
eccentricities suggest that
  these planets have experienced damping, as suggested also by the
resonant repulsion theory (\citealp{lithwua}, see also
\citealp{baty}).  In that work, we 
 found that if Kepler planets have experienced
substantial energy dissipation, but substantially less angular
momentum damping, the two planets will be repelled from each
other.  This would explain the observed
pile-up of planets just wide of resonances. A corollary of this theory
is that low mass planets in the Kepler sample should have little if
any free eccentricity.

Although this appears to be confirmed by three of the six systems we
analyze, we are puzzled by the small but finite free eccentricities in
the other three systems.    Assuming that resonant repulsion indeed occurred, it
   would require that the planets' eccentricities
  were subsequently excited, perhaps by interplanetary
  interactions. Such a scenario would argue against tides as the
  mechanism of dissipation causing resonant repulsion, because tides would
  have to act over very long times to be effective.  Instead,
  dissipation by a disk of gas or planetesimals are more plausible
  damping agents.

  TTV data have also been reported for Kepler 9b/c and Kepler 30b/c
  \citep{HolmanKepler9,FabryckyIV}. These pairs contain one or two
  giant planets. We find preliminary evidence that these systems have
  large TTV phases. This, if true, will indicate the presence of large
  free eccentricities in systems of giant planets, in contrast to the
  lower mass planets discussed here. We are currently analyzing Kepler
  public lightcurve data to distill more TTV systems. One interesting
  issue to pursue is whether planet pairs at much shorter or much
  longer orbital periods have the same characteristics as those
  analyzed here.

  Many planet pairs are also near resonance with a third planet. This
  may bring further complications to our TTV analysis but is not
  considered here.

\subsection{ Values of Planet Masses}

All six systems we analyze have orbital periods between 6 and 15 days,
and planet radii ranging from $2$ to $7 R_\oplus$.  We confirm the
mass estimates of \citet{CochranKepler18} for Kepler 18c/d, and the
mass estimates of \citet{FabryckyIV} for Kepler 25b/c, under the assumption
that these systems have $|\E|\lesssim|\Delta|$, consistent with their small phase. Densities of
these planets range between $0.3$ and $2\g/\cm^3$.  For Kepler 28b/c,
32b/c, we obtain nominal mass upper limits that lead to density upper limits
of $\sim 3\g/\cm^3$. We also argue that these upper limits
  are likely not too different from the real masses.
For Kepler 24b/c and 23b/c, the nominal densities are $\sim
10\g/\cm^3$.

\acknowledgements

 Y.L. acknowledges support from NSF grant
AST-1109776.
 JWX and YW acknowledge support by NSERC and the Ontario
government.
We thank the referee, Dan Fabrycky, for a helpful report.

\bibliographystyle{apj}
\bibliography{orig}

\clearpage
\appendix
 \section{Derivation of analytical TTV}
 \label{sec:derive}
 
  We derive the transit time variations for two planets near a first order ($j\!:\!\!j\!-\!1$) mean motion
 resonance, assuming that the planets are coplanar with each other and with the line of sight.
 We  solve perturbatively, taking the following quantities to be small:
 the eccentricities ($e,e'$), the mass ratios of the planets to the star ($\mu,\mu'$) and the fractional distance
 to resonance ($\Delta$).  Typical values for Kepler planets are, very roughly 
 $e\lesssim 0.1$, $\mu\lesssim 10^{-4}$, and $|\Delta|\lesssim 0.05$. 
 Section \ref{sec:assump} provides further restrictions on these parameters for our perturbative
 treatment to be valid.

 We first give a brief overview of the calculation that follows.
 When the equations of motion are solved perturbatively (where the unperturbed solution
 is  circular), the leading order solution for the complex eccentricity is a sum of
 forced and free terms (Eq. \ref{eq:zans}); the leading order solution for the semimajor axis is
 determined by  the product of the free and forced eccentricities (Eq. \ref{eq:da}); and 
 the perturbed solution for the longitudes is determined by the perturbed semimajor axis (divided by $\Delta$; 
 Eq. \ref{eq:lamprime}).   These solutions determine
  the angular positions of the planets as functions of time (via Eq. \ref{eq:thlam}).
Finally, the angular positions are trivially inverted to obtain the times of transit, and hence the TTV.

   The total energy, or Hamiltonian, is \citep[e.g.,][]{MD00}
 \beqn H=-{GM_*m\over 2a}-{GM_*m'\over 2a'} - {Gmm'\over a'}R^j \ ,
\label{eq:res}
\eeqn
where $M_*$ is the stellar mass, and
the disturbing function due to the $j:j-1$ resonance is
\beqn
R^j= f e\cos(\lambda^j-\pomega)
+g e'\cos(\lambda^j-\pomega')  \ ,
\label{eq:bc}
\eeqn
for
\beqn
\lambda^j\equiv j\lambda'-(j-1)\lambda \ .
\eeqn
In the above, $\{m,a,e,\lambda,\pomega,m',a',e',\lambda',\pomega' \}$ are
the mass and standard orbital elements for the inner (unprimed) and outer (primed) planets,
following the notation of \cite{MD00}.
The coefficients $f$ and $g$
are order-unity, and are functions of 
$j$ and
\be
\alpha\equiv {a/a'}\  .
\ee
These are tabulated for a few resonances in  Table \ref{tab:simp}

The equations of motion are  Hamilton's equations, 
after expressing the orbital elements in the above Hamiltonian in terms 
of canonical Poincar\'e variables \citep[e.g.][]{MD00}.
 The resulting equation for the inner planet's longitude is
\beqn
{d\lambda\over dt}&=&{1\over\sqrt{GM_*}}{2\sqrt{a}\over m}{\partial H\over\partial a}   \\
&\approx& \sqrt{{GM_*\over a^3}} \label{eq:lam} \ ,
\eeqn
and similarly for $\lambda'$.
We drop the derivative of the disturbing function in the above, and
justify this below.
Variations in the semimajor axes are second order in eccentricity
   (see below); hence
\be
\left(\begin{array}{c}{\lambda} \\ \lambda'\end{array}\right)=
\left(\begin{array}{c}
{2\pi\over P}(t-T) \\ 
{2\pi\over P'}(t-T')
 \end{array}\right) 
 +\left(\begin{array}{c}{\delta\lambda} \\ \delta\lambda'\end{array}\right)
 \label{eq:lambar}
\ee
where the periods are
\beqn
P\equiv2\pi\sqrt{a^3\over GM_*}  ,\ 
P'\equiv2\pi \sqrt{a'^3\over GM_*} \ .
\eeqn
The periods will henceforth be treated as constants (i.e., considered to be functions
of the unperturbed semimajor axes);
$T$ and $T'$ are constant
reference times; and the variations in the longitudes 
 $\delta\lambda,\delta\lambda'=O(e^2)$ remain to be determined.

For the eccentricity equations, we introduce the complex eccentricities
 \citep[e.g.,][]{ogilvie}
\beqn
z&\equiv& ee^{i\pomega} \\
z'&\equiv& e'e^{i\pomega'} \ , 
\eeqn
in terms of which the disturbing function may be written as
\beqn
R^j={1\over 2}(fz^*+gz'^{*})e^{i\lambda^j} + {\rm c.c.} \ ,
\eeqn
where c.c. denotes the complex conjugate of the preceding term.
The eccentricity equation for the inner planet is
\beqn
{dz\over dt}&=&-{1\over\sqrt{GM_*}}{2i\over m\sqrt{a}}{\partial H\over \partial z^*} 
\eeqn
to leading order in eccentricity, and similarly for the outer planet, i.e.
\beqn
{d\over dt}\left(\begin{array}{c}z \\z'\end{array}\right)=
i{2\pi\over P'}\left(\begin{array}{c}\mu' f/\sqrt{\alpha} \\ 
\mu g
\end{array}\right)e^{i\lambda^j}
\eeqn
where the mass ratios are
\be
\mu\equiv m/M_* , \  \mu'\equiv m'/M_*.
\ee
Solving the eccentricity equations to first order yields
a sum of free and forced terms:
\beqn
\left(\begin{array}{c}z \\ z'\end{array}\right)=
\left(\begin{array}{c}z_{\rm free} \\z'_{\rm free}\end{array}\right)
-{1\over j\Delta}\left(\begin{array}{c}\mu'f/\sqrt{\alpha} \\ 
\mu g
\end{array}\right)e^{i\lambda^j} \ ,
\label{eq:zans}
\eeqn
where the free terms are constant
and the normalized distance to resonance is
\be
\Delta\equiv {j-1\over j}{P'\over P}-1 \ , \label{eq:DeltaX}
\ee
assumed to satisfy $|\Delta|\ll 1$.
In the above, we have used the relation
\be
{d\over dt}\lambda^j=-(j\Delta) {2\pi\over P'} + O(e^2)
\ee

We proceed to determine the $O(e^2)$ changes to $a,a'$, and thereby to
obtain $\delta\lambda,\delta\lambda'$ to this order.  To do so, we
note that the resonant Hamiltonian (Equation (\ref{eq:res})) has two
constants of motion in addition to the energy:
$K=\Lambda+(j-1)(\Gamma+\Gamma')$ and $K'=\Lambda'-j(\Gamma+\Gamma')$,
where the $\Lambda$ and $\Gamma$ are the usual Poincar\'e momenta
\citep{MD00}.  From the constancy of $K$, the variation in $a$ over
the course of the planets' orbits (i.e. $\delta a$) satisfies \beqn
m\sqrt{a}{\delta a\over 2a}+(j-1) \left( m\sqrt{a}{e^2\over 2}+
  m'\sqrt{a'}{e'^2\over 2} \right)={\rm const} \ , \eeqn discarding
terms of higher order in $\delta a,\delta a',e^2,e'^2$.  Inserting the
eccentricities from Equation (\ref{eq:zans}), only the cross terms
between the free and forced eccentricities yield time-varying
components to $\delta a$, implying
\beqn
\left(\begin{array}{c} \delta a/a\\
\delta a'/a'
\end{array}\right)
=
\left(\begin{array}{c}{j-1\over j}\mu'/ \sqrt{\alpha} \\
 -\mu\end{array}\right){\E^*\over\Delta}e^{i\lambda^j}+c.c.
 \label{eq:da}
\eeqn
where
\be
\E\equiv  fz_{\rm free}+gz'_{\rm free} 
\label{eq:zfree}
\ee
is a weighted sum of the two planets' free eccentricities.

The equation for the longitudes (Equation (\ref{eq:lam}))
 becomes, to first order in $\delta a$,
\be
{d\over dt}\delta\lambda =-{3\over 2}{2\pi\over P}{\delta a\over a}  \ ,
\label{eq:lamdot}
\ee
and similarly for the outer planet.
The solutions are
\beqn
\left(\begin{array}{c}\delta\lambda \\ \delta\lambda'\end{array}\right)=
\left(\begin{array}{c}
 \mu'{j-1\over j}\alpha^{-2}
 \\ 
 -\mu
 \end{array}\right){3\E^*\over 2i j\Delta^2}e^{i\lambda^j}+c.c. 
 \label{eq:lamprime}
 \ ,
\eeqn

Now, to convert from $\lambda$ to $\theta$, we must add the following, 
valid to first order in eccentricity,
\beqn
\theta-\lambda&=&2e\sin(\lambda-\pomega) 
\label{eq:thlam}
\\
&=&{z^*\over i}e^{i\lambda}+c.c. 
\eeqn
and similarly for primed quantities.
Since we are ultimately interested in transit times, we need only
 insert for $z$ the forced eccentricity  (Equation (\ref{eq:zans})), 
because the free eccentricity produces a term
with the same period as the transits, and hence does produce variations
from transit to transit.
By the same logic, we may 
 drop the $e^{i\lambda}$ multiplying the forced eccentricity
 if we choose the observer to be at angular position 0. We then have
\beqn
\left(\begin{array}{c}\theta-\lambda \\ \theta'-\lambda'
\end{array}\right)=
{1\over j\Delta}
\left(\begin{array}{c}\mu' f/\sqrt{\alpha} \\ 
\mu g
\end{array}\right){e^{i\lambda^j}\over i}+c.c.
\label{eq:lamcon2}
\eeqn
The final expression for $\theta,\theta'$  is given by 
the sum of Equation (\ref{eq:lambar}) (after inserting Equation (\ref{eq:lamprime})) with 
Equation (\ref{eq:lamcon2}), yielding
\beqn
\theta &=& {2\pi\over P}(t-T)-{2\pi\over P}\{{V\over 2i}e^{i\lambda^j}+c.c. \} \label{eq:thf} \\
\theta' &=& {2\pi\over P'}(t-T')-{2\pi\over P'}\{{V'\over 2i}e^{i\lambda^j}+c.c. \} \label{eq:thfp}
\eeqn
where the  amplitudes are
\beqn
V&=&{P\over \pi}{\mu'\over j\Delta}\alpha^{-1/2}
\left({-f-{j-1\over j}\alpha^{-3/2}{3\E^*\over 2\Delta}}  \right)
\label{eq:vap}
\\
V'&=&{P'\over\pi}{\mu\over j\Delta}\left(
- g+{3\E^*\over 2\Delta}
\right) \ .
\label{eq:vpap}
\eeqn
Note that $\alpha=\left((1+\Delta){j\over j-1}\right)^{-2/3}$, whence follows Equation (\ref{eq:v}) in the
body of the paper   after dropping $O(\Delta)$ corrections.  (We somewhat
inconsistently keep
 the $O(\Delta)$ corrections to $f$ and $g$ in Table \ref{tab:simp} because these
 can be quite large, especially near the 4:3 and 5:4 resonances).
Since we choose $\theta=0$ to point along the line of sight, 
transits of the inner planet occur whenever $\theta/2\pi=0,1,\cdots$, and similarly
for the outer planet. 
We conclude that the TTV signals for the inner and outer planets are as given in
Equations (\ref{eq:tv})--(\ref{eq:zfreeloc}).

\begin{table}[t]
  \caption{Coefficients of the disturbing function, defined via Equations (\ref{eq:res})--(\ref{eq:bc}).
    Numerical values   expanded to first order in $\Delta$,
    the relative distance to resonance (Eq. \ref{eq:DeltaX}).  
  }  
\begin{minipage}{160mm}
\begin{tabular}{|c|ccccc|}
\hline
 &             $j:j-1$\footnote{
  $b_s^j\equiv b_s^j(\alpha)$ are Laplace coefficients and  $D\equiv d/d\alpha$
\citep{MD00}.}
                &               $2:1$          &                $3:2$         &         $4:3$  
 &            $5:4$                
 \\ \hline\hline
     $f$    &
     $-jb_{1/2}^j-{\alpha\over 2}Db_{1/2}^j$    & 
     $-1.190+2.20\Delta$ & 
     $-2.025+6.21\Delta$ &
     $-2.840+12.20\Delta$  &
      $-3.650+20.15\Delta$
       \\
     $g$  &  
     $(j-{1\over 2}) b_{1/2}^{j-1}+{\alpha\over 2} Db_{1/2}^{j-1}+{\rm indirect}_{2:1}$  &
     $0.4284-3.69\Delta$\footnote{
     The 2:1 value of $g$
contains the indirect term for an internal perturber  ($=-1/(2\alpha^2)$).  One should use
this  in Equations (\ref{eq:vp})--(\ref{eq:zfreeloc}) for the 2:1;
however, in Equation (\ref{eq:v}) one should use $\E=fz_{\rm free}+g_{\rm ext}z_{\rm free}'$, where $g_{\rm ext}=0.4284-1.17\Delta$, appropriate for an external perturber. Nonetheless, this distinction 
is unlikely to be of practical importance unless one is interested
in the $O(\Delta)$ corrections to the free eccentricities.}
     &
      $2.484-5.99\Delta$   & 
      $3.283-11.9\Delta$ &
       $4.084-19.86\Delta$ 
       \\ 
       \hline
\end{tabular}
\end{minipage}
\label{tab:simp}
\end{table}

\section{Assumptions and Range of Validity}
\label{sec:assump}

\begin{itemize}

\item We have assumed the system is not locked in resonance.   Resonance locking occurs
when $\delta\lambda\sim$ unity.  From Equation (\ref{eq:lamprime}), this implies
that our expressions are valid as long as
\be
 e_{\rm free} \lesssim \Delta^2/\mu \ . \nonumber
\ee 
For typical Kepler systems, $|\Delta|\gtrsim 1\%$ and $\mu\lesssim 10^{-4}$, and hence this assumption is usually valid. 

\item  In Equation (\ref{eq:lam}), we dropped the derivative of the
disturbing function.   This term is $\sim \mu e$, and hence is smaller
than the term that is kept by $\sim \Delta$ (Eq. \ref{eq:lamdot}).

\item We have neglected secular effects.  Secular precession occurs on
  the timescale $\sim P/\mu$, which is much longer than the timescale
  of the resonant effects considered here, $\sim P/\Delta$.  Hence
  secular effects only lead to corrections to our formulae of order
  $|\mu/\Delta|\ll 1$. Nonetheless, secular precession causes $\E$ to
  precess on thousand-year timescales, which randomizes the
  orientation of $\E$.  For that reason, we argue in this paper that
  if  the data requires $\angle\E=0$ (or $\pi$) for many systems, it
  suggests that their $\E$ nearly vanishes.

\item The timescale for general relativistic precession is also too
  long to be important, specifically $(d\omega/dt)^{-1}\approx
  Pac^2/(6\pi GM_*)\approx 4000\times P_{\rm 5\ day}^{5/3} {\rm
    year/radian}$.

\item We assume perfect coplanarity. The small
    inclination dispersion inferred for Kepler multi-planet systems
justifies this assumption.


\end{itemize}

\end{document}